# OPTICAL PROPERTIES AND STRUCTURE OF MOST STABLE SUBNANOMETER $(ZnAs_2)_n$ CLUSTERS

O.A. Yeshchenko[*], I.M. Dmitruk, S.V. Koryakov, M.P. Galak, I.P. Pundyk, L.M. Hohlova

*Physics Department, National Taras Shevchenko Kyiv University,*

*2/1 Akademik Glushkov prosp., 03127 Kyiv, Ukraine*

$ZnAs_2$ nanoclusters were fabricated by incorporation into pores of zeolite Na-X and by laser ablation. Absorption and photoluminescence spectra of $ZnAs_2$ nanoclusters in zeolite were measured at the temperatures of 4.2, 77 and 293 K. Both absorption and PL spectra consist of two bands which demonstrate the blue shift from the line of free exciton in bulk crystal. We performed the calculations aimed to find the most stable clusters in the size region up to size of the zeolite Na-X supercage. The most stable clusters are $(ZnAs_2)_6$ and $(ZnAs_2)_8$ with binding energies of 7.181 eV and 8.012 eV per $(ZnAs_2)_1$ formula unit respectively. Therefore, we attributed two bands observed in absorption and PL spectra to these stable clusters. The measured Raman spectrum of $ZnAs_2$ clusters in zeolite was explained to be originated from $(ZnAs_2)_6$ and $(ZnAs_2)_8$ clusters as well. The PL spectrum of $ZnAs_2$ clusters produced by laser ablation consists of a single band which has been attributed to emission of $(ZnAs_2)_8$ cluster .

## 1. Introduction

Many different methods have been used for fabrication of the semiconductor nanoparticles, e.g. fabrication of nanoparticles in solutions [1], glasses [2] or polymers [3]. However, it is not easy to control the size distribution of small nanoparticles with countable number of atoms (so called clusters) in these methods. Matrix method based on the incorporation of materials into the 3D regular system of voids and channels of zeolites crystals could be one of the possible solutions [4,5]. Moreover, the subnanometer and nanometer clusters are the very interesting as they are in intermediate position between the molecules and the typical nanocrystals. Usually, the structure of nanoclusters is different from the structure of nanocrystals which resembles the structure of bulk crystals [6]. As a rule the methods of calculation of the structure of electronic states of nanocrystals which are based on the effective mass approximation are not applicable for clusters. Thus, nanoclusters are the very interesting objects as their structure, electronic and vibrational properties

---

[*] Corresponding author: O.A.Yeshchenko.

Tel.: +380-44-5264587;   Fax: +380-44-5264036; E-mail: yes@univ.kiev.ua

are quite different from the crystalline nanoparticles. Zeolites provide the opportunity to obtain extremely small clusters in the pores with diameters up to 15 Å. Zeolites are crystalline alumosilicates with cavities which diameter can vary in the range from 7 to 15 Å. It depends on the type of alumosilicate framework, ratio Si/Al, origin of ion-exchanged cations, which stabilise negative charge of framework, etc. Zeolite Na-X, which has been used in the present work has Si/Al ratio equal 1, Fd3m symmetry and two types of cages: one is sodalite cage – truncated octahedron with diameter 8 Å and supercage, which is formed by the connection of sodalites in diamond-like structure with the diameter of about 13 Å [7]. All cages are interconnected by shared small windows and arranged regularly. Thus, the cages can be used for fabrication of small semiconductor nanoclusters.

Laser ablation (LA) is a well-known method to produce nanoclusters by ablating material from a solid target [8]. LA usually is performed in vacuum, or sometimes in inert gas such as Ar or more reactive gases such as ammonia or nitrogen. Recently a new variation of LA has been reported whereby the target is immersed in a liquid medium, and the laser beam is focused through the liquid onto the target surface [9]. As a rule, the nanoclusters formed at the ablation have diameters from several angstroms to several tens angstroms. LA technique has been used to produce nanoclusters of semiconductors (see e.g. Refs. [10,11]) and metals (see e.g. Ref. [12]).

The nanoclusters of II-V semiconductors are studied rather poorly. To our knowledge there are several works on $Cd_3P_2$ nanoclusters fabricated by wet chemistry methods [13,14] and by thermolysis [15] and alcoholysis of organometallic species [16]. As well, in our recent work [17] we have reported the fabrication and study of the optical properties of the nanoclusters of another II-V semiconductor ($ZnP_2$) incorporated into zeolite Na-X matrix. The present paper is the first study of the nanoclusters of another II-V semiconductor: zinc diarsenide ($ZnAs_2$). Wet chemistry methods seem to us not to be suitable for production of ultrasmall II-V nanoclusters due to their high reactivity in water. It is hard to expect their high stability in glass melt as well. Thus, incorporation into zeolite cages and production by laser ablation seem to us to be ones of the most suitable methods of fabrication of II-V semiconductor nanoclusters.

Quantum confinement of charge carriers in nanoclusters leads to new effects in their optical properties. Those are the blue shift of exciton spectral lines originating from the increase of the kinetic energy of charge carriers and the increase of the oscillator strength per unit volume [18,19]. These effects are quite remarkable when the radius of the nanoparticle is comparable with or smaller than Bohr radius of exciton in bulk crystal. Incorporation into zeolite pores and laser ablation are quite promising methods to produce small nanoclusters in which these effects are considerable to be studied.

Bulk ZnAs$_2$ crystal is the direct-gap semiconductor. The symmetry of its lattice is characterised by the space symmetry group $C_{2h}^5$ (monoclinic syngony). As this biaxial crystal is strongly anisotropic, its optical spectra are characterised by three exciton series. Since the lowest energy exciton peak is observed at 1.0384 eV [20,21], the blue shifted exciton lines of ZnAs$_2$ nanoclusters are expected to be in the visible spectral region.

## 2. Structure and optical properties of ZnAs$_2$ nanoclusters in zeolite Na-X matrix

For the fabrication of ZnAs$_2$ nanoclusters we used ZnAs$_2$ bulk crystals and synthetic zeolite of Na-X type. The framework of zeolite Na-X consists of sodalite cages and supercages with the inner diameters of 8 and 13 Å, respectively. ZnAs$_2$ nanoclusters probably are too large to be incorporated into small sodalite cage, because of the existence of many Na cations. Therefore, it is naturally to assume that only the supercages can be the hosts for the nanoclusters. Zeolite and ZnAs$_2$ crystals were dehydrated in quartz ampoule in vacuum about $2\times10^{-5}$ mm Hg for 1 h at 400°C. Then ampoule was sealed. We used 100 mm length ampoule for space separation of semiconductor source and zeolite in it. The fabrication of samples was carried out in two stages. At the first stage (see Fig.1) ZnAs$_2$ was incorporated into the zeolite matrix through the vapour phase at 776°C in source region and 770°C in zeolite region for 100 h. At the second stage, the inverted temperature gradient was applied: 751°C in source region and 767°C in zeolite region. The duration of the second stage was 40 h. The cooling of ampoule we carried out gradually with above mentioned inverted temperature gradient. The stability of structure of lattice of zeolite single crystals was controlled by XRD method. The control showed that at above mentioned temperatures the zeolite lattice structure was stable, i.e. semiconductor nanoclusters were incorporated into the single crystal zeolite matrix.

During the optical measurements the samples were in vacuum in quartz ampoule. A tungsten-halogen incandescent lamp was used as a light source for the diffuse reflection measurements. An Ar$^+$ laser with wavelength 488.0 nm was used for the excitation of the luminescence of ZnAs$_2$ nanoclusters. The absorption spectra of the nanoclusters were obtained from the diffuse reflection spectra by conversion with Kubelka-Munk function $K(\hbar\omega)=[1-R(\hbar\omega)]^2/2R(\hbar\omega)$, where $R(\hbar\omega)$ is the diffuse reflectance normalised by unity at the region of no absorption.

Diffuse reflection (DR) and photoluminescence (PL) spectra of the ZnAs$_2$ nanoclusters incorporated into the 13 Å supercages of zeolite Na-X were measured at room (290 K), liquid nitrogen (77 K) and liquid helium (4.2 K) temperatures. Then, the DR spectrum was converted to absorption one by the Kubelka-Munk method described above. Within the accuracy of determination of bands spectral positions we did not observe noticeable change of both absorption and PL spectra with temperature. The absorption spectrum obtained by Kubelka-Munk method is presented on Fig. 2(*a*). The spectrum demonstrates clear two-band structure. The spectral positions of the respective bands signed as $B_1$ and $B_2$ are presented in Table 1. Both the bands demonstrate blue shift (Table 1) from the exciton line of the bulk crystal. Let us note that the blue shift of the high-energy absorption band for $ZnAs_2$ clusters in Na-X zeolite has value close to the respective one for $ZnP_2$ clusters in the same zeolite [17]: 0.881 eV for $ZnAs_2$ and 0.808 eV for $ZnP_2$. The observed blue shift allows us to attribute these bands to the absorption into the first electronic excited state of $ZnAs_2$ nanoclusters incorporated into supercages of zeolite. The photoluminescence spectrum of $ZnAs_2$ clusters in zeolite (Fig. 2(b): solid curve) shows the same structure as the absorption one, i.e. PL spectrum consists of the corresponding two $B'_1$ and $B''_1$ bands. Their spectral positions are presented in Table 1. PL bands of nanoclusters are blue shifted from the spectra of bulk crystal as well. The observed blue shift of the absorption and luminescence bands is the result of the quantum confinement of electrons and holes in $ZnAs_2$ nanoclusters. As the exciton Bohr radius in bulk crystal is larger than radius of zeolite supercage, the strong confinement regime takes place in the nanoclusters.

It is often observed that nanoclusters with certain number of atoms are characterised by elevated stability (ultrastable nanoclusters) and are more abundant in the sample. This effect is well known for the nanoclusters of different types, e.g. for C [22], Ar [23], Na [24], and for nanoclusters of II-VI semiconductors [25, 26]. Our first-principles calculations [17] have shown that such stable nanoclusters exist for $ZnP_2$. Those are $(ZnP_2)_6$ and $(ZnP_2)_8$ with binding energies 5.16 eV and 6.42 eV per formula unit respectively. Thus, it would be naturally to assume that similarly to $ZnP_2$ the respective stable $(ZnAs_2)_n$ nanoclusters exists for $ZnAs_2$ as well. We performed the calculations aimed to find such stable $ZnAs_2$ clusters. Initially, we performed the geometry optimization of the structure of clusters by molecular mechanics method. Then, we performed the *ab initio* calculation of the ground state energy of the clusters with optimized structure. The results are presented in Fig.3. It is seen from the figure that $ZnAs_2$ molecule does not exist as it is unbound. The clusters with even *n* have higher binding energy than clusters with odd n. Likely to $ZnP_2$ the $(ZnAs_2)_n$ clusters with $n = 6$ and 8 are the most stable. The $(ZnAs_2)_6$ cluster is characterised by the binding energy 7.18 eV per formula unit, and the $(ZnAs_2)_8$ one – 8.01 eV per formula unit, i.e. the $(ZnAs_2)_n$ clusters are stronger bound than the respective $(ZnP_2)_n$ ones. The maximum diameter of $(ZnAs_2)_6$ cluster is 8.52 Å, and the maximum diameter of $(ZnAs_2)_8$ is 9.28 Å. Here and everywhere in the article, maximum diameter means the distance between the centers of outermost atoms of cluster. The structure of these ultrastable clusters is presented in Fig.4. It is seen that as it can be expected the structure of $(ZnAs_2)_6$ and $(ZnAs_2)_8$ clusters is the same as the structure of the respective $(ZnP_2)_n$ clusters. $(ZnAs_2)_6$ looks like some nanotube, and $(ZnAs_2)_8$ looks like some nanodisk. Similar to the structure of bulk $ZnAs_2$ crystal [26] both these clusters have six-membered rings of atoms. Most probably, some other stable $(ZnAs_2)_n$ clusters exist as well. One of those is $(ZnAs_2)_{10}$ cluster with maximum diameter of 10.47 Å and binding energy of 7.89 eV per formula unit. However, our estimation of the diameter of largest clusters that might be placed in zeolite Na-X supercage is 9.68 Å. Our estimations consider the van der Waals radii of Zn (1.274 Å) and As (2.050 Å). Therefore, the clusters with $n > 8$ can not be incorporated into supercages of Na-X zeolite. Since the $(ZnAs_2)_6$ and $(ZnAs_2)_8$ clusters are the most stable, it is quite reasonable to assume that these clusters are the most abundant. Thus, $B_1$ and $B_2$ bands can be attributed to absorption into the first excited state of above mentioned stable $(ZnAs_2)_6$ and $(ZnAs_2)_8$ nanoclusters incorporated into the supercages of zeolite matrix.

The observed blue shift of the absorption and luminescence bands is the result of the quantum confinement of electrons and holes in $ZnAs_2$ particles. The exciton Bohr radius $a$ in bulk crystal is 21.6 Å [20] that is larger than radius of zeolite supercage (6.5 Å). So, the strong confinement model can be used to consider the quantum confinement of the carriers in nanoparticles. Taking the diameters of $(ZnAs_2)_6$ and $(ZnAs_2)_8$ clusters as 8.52 Å and 9.28 Å correspondingly, one can estimate the values of blue shift for these clusters by the theory developed in Ref.[27]. The estimations give the following values of blue shift: 3.254 eV and 2.717 eV for $(ZnAs_2)_6$ and $(ZnAs_2)_8$ clusters respectively. The experimental values of blue shift for corresponding $B_1$ and $B_2$ absorption bands are 0.881 eV and 0.626 eV correspondingly. Thus, the value of blue shifts calculated by the effective mass approximation are substantially different from the experimental ones. It means probably that the effective mass approximation fails for clusters with small number of atoms. Another explanation of the obtained small value of the blue shift is the shift of electron and hole energy levels to the low energy due to the tunnelling between the neibouring supercages.

One can see from the Table 1 that the luminescence bands have the Stokes shift from the absorption ones. The values of this shift are large enough: 0.290 eV for $B_1'$ band and 0.355 eV for $B_2'$ one. These values are considerably larger than ones for $ZnP_2$ nanoclusters in the same Na-X zeolite (0.078-0.135 eV: see Ref. [17]). Stokes shift is well known both in the molecular spectroscopy and in the spectroscopy of nanoclusters. It is known that this kind of Stokes shift (so-called Frank-Condon shift) is due to vibrational relaxation of the excited molecule or nanoparticle to the ground state. The theory of Frank-Condon shift in nanoclusters was developed in Ref. [28] where the first-principle calculations of excited-state relaxations in nanoclusters were performed. As it is shown in Ref. [28], for small nanoclusters the Stokes shift is the Frank-Condon one, which is the result of the vibrational relaxation of the nanoparticle in the excited electronic state. The considerable values of Stokes shift in $(ZnAs_2)_n$ clusters mean the substantial role of vibrational relaxation in excited nanoparticles.

To check our above assumption of the origin of the absorption and PL bands of the $ZnAs_2$ incorporated into zeolite we performed Raman study of the samples. The obtained Raman spectrum is shown in Fig.5. It is seen that the intensity of the spectrum is quite low. Probably, this is due to small relative portion of the scattering $ZnAs_2$ clusters with respect to total amount of zeolite in samples. The Raman spectrum consists of a single rather wide band centered at 250 cm$^{-1}$ with half-width of 82 cm$^{-1}$. To check our above assumption of the formation of (ZnAs$_2$)$_6$ and (ZnAs$_2$)$_8$ clusters in zeolite supercages we performed semi-empirical calculation of the frequencies of vibrations of these clusters. Calculations showed that (ZnAs$_2$)$_6$ cluster would have the Raman active group of vibrational normal modes with frequencies in the range 221.0 – 308.5 cm$^{-1}$, and (ZnAs$_2$)$_8$ one – in the range of 167.6 – 282.7 cm$^{-1}$. Further, we calculated the total theoretical Raman spectrum where the contributions of both the clusters were added. The theoretical spectrum was constructed as a sum of gaussians centered at the corresponding frequencies. It is seen from Fig.5 that theoretical spectrum correspond with experimental one quite good. Therefore, the Raman spectrum proves our assumption of the formation of (ZnAs$_2$)$_6$ and (ZnAs$_2$)$_8$ clusters in zeolite pores.

### 3. Photoluminescence of ZnAs$_2$ nanoclusters produced by laser ablation method

For ablation the pulsed Cu laser ($\lambda$=5782 Å) was used. The pulse intensity of the focused laser beam was about 1.5 MW/cm$^2$, pulse duration was of 20 ns at a repetition rate of 10 kHz. The beam was focused on the surface of target to a spot-size diameter of approximately 0.5 mm. During the ablation the target (bulk ZnAs$_2$ crystal of area 10 mm$^2$) was dipped into the liquid nitrogen. The produced by ablation nanoclusters of ZnAs$_2$ were deposited on quartz plate that was positioned on the distance of about 0.2 mm from the irradiated surface of the target crystal. The run time of ablation was 20 min. The photoluminescence spectrum was measured from the deposited film by cw Ar$^+$ laser ($\lambda$=488.0 nm).

The obtained PL spectrum of ZnAs$_2$ clusters is shown in Fig.3(b). The spectrum consists of a single band marked as $B_2''$. It is shown from the figure and Table 1 that the spectral position of its maximum coincides with the position of $B_2'$ band of luminescence spectrum of ZnAs$_2$ clusters in zeolite. This fact and the proximity of values of the half-widths of $B_2''$ and $B_2'$ bands (0.13 and 0.18 eV correspondingly) allows us to assume the same origin of these PL bands, i.e. that both the bands originate from the emission from excited to ground electronic state of (ZnAs$_2$)$_8$ cluster. This our assumption seems to be quite reasonable as such cluster is the most stable. Correspondingly, (ZnAs$_2$)$_8$ clusters would be formed in prevalent quantities at the ablation. An effect of the prevalence of the most stable nanoclusters in mass spectra is well known for II-VI semiconductors (see e.g. Refs. [24,25]). Meanwhile, laser ablation is the method for production of analysed particles in mass spectrometry. It is naturally that other ZnAs$_2$ nanoclusters besides (ZnAs$_2$)$_8$ would be formed at the ablation as well, but, probably, their quantity is quite small compared to quantity of (ZnAs$_2$)$_8$ ones. Therefore, these less abundant clusters would not give the considerable contribution to PL spectrum.

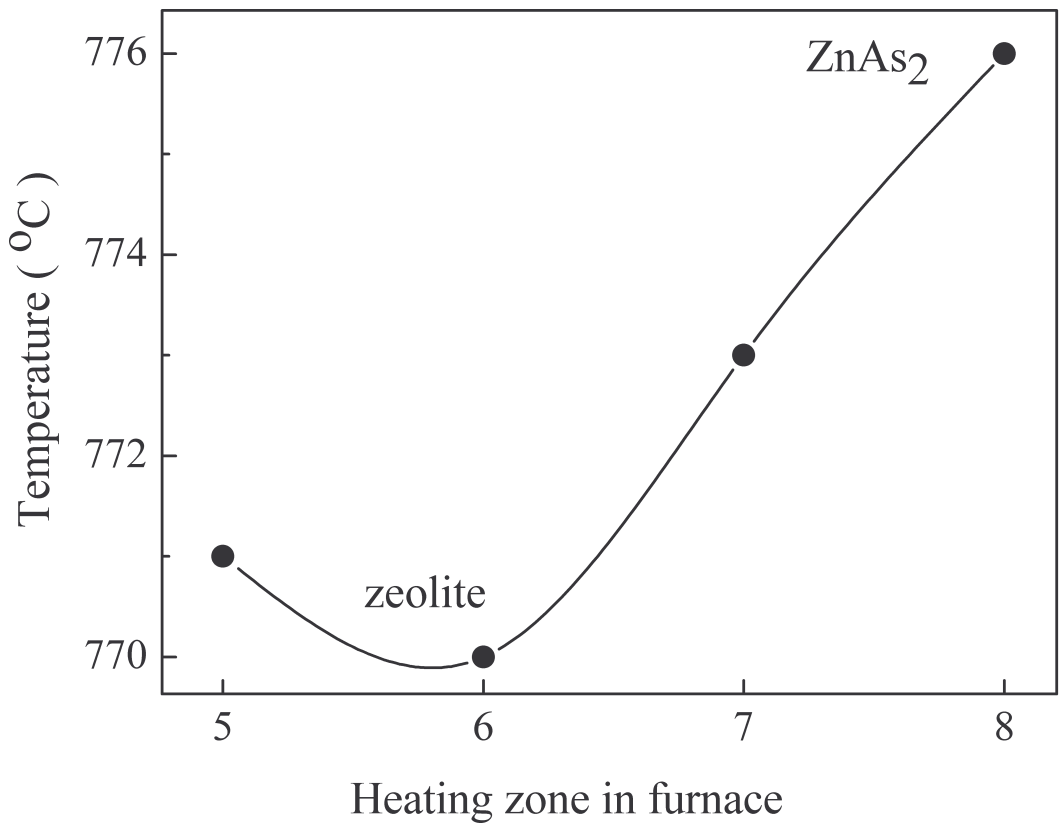

Fig.1.The temperature gradient used at first stage of the fabrication of $ZnAs_2$ clusters in zeolite Na-X.

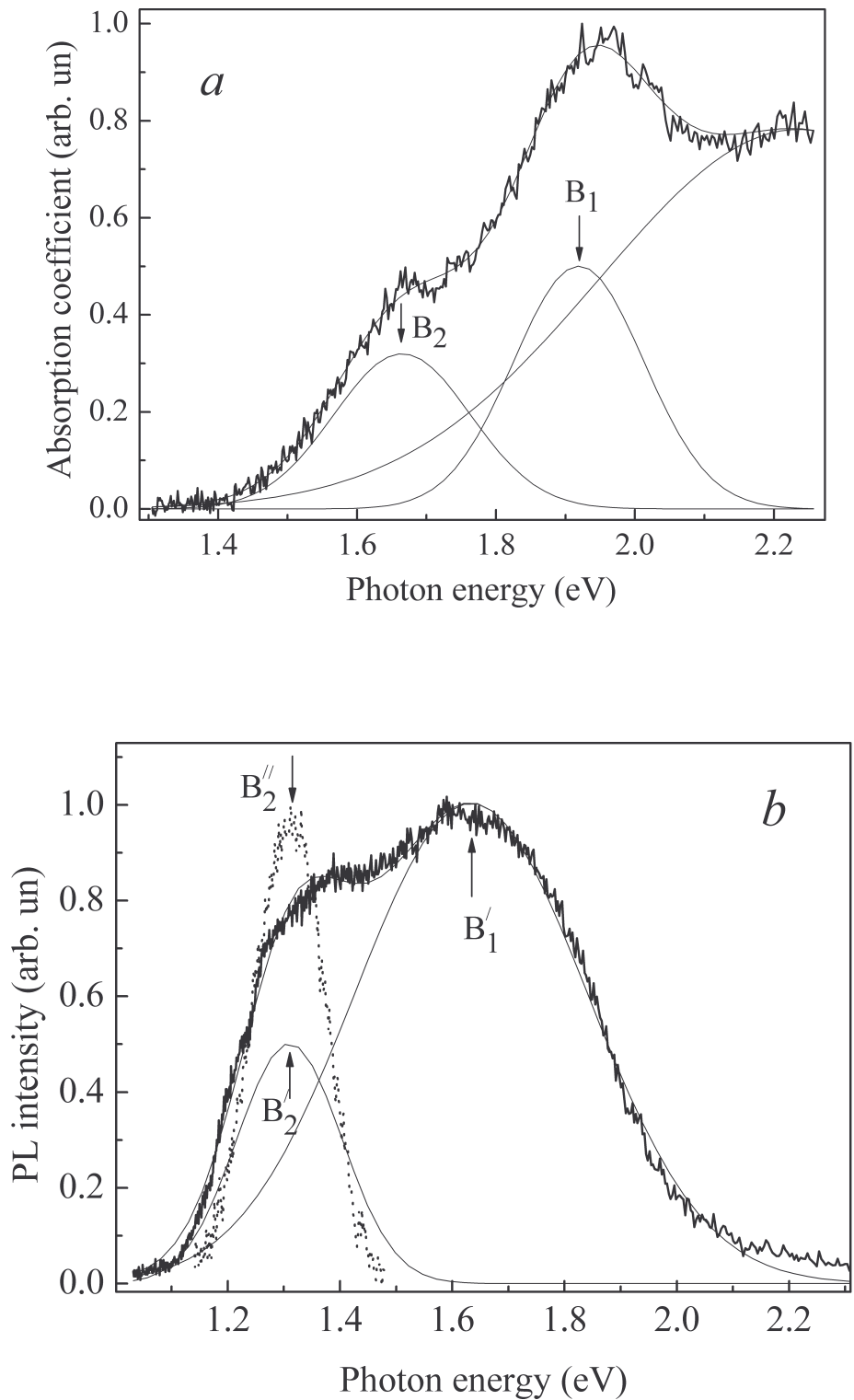

Fig. 2. (a) – The obtained by Kubelka-Munk method absorption spectrum of $ZnAs_2$ clusters in zeolite Na-X at the temperature of 77 K. (b) – The photoluminescence spectrum of $ZnAs_2$ clusters in zeolite Na-X (solid line) and clusters produced by laser ablation (dotted line) at the temperature of 77 K.

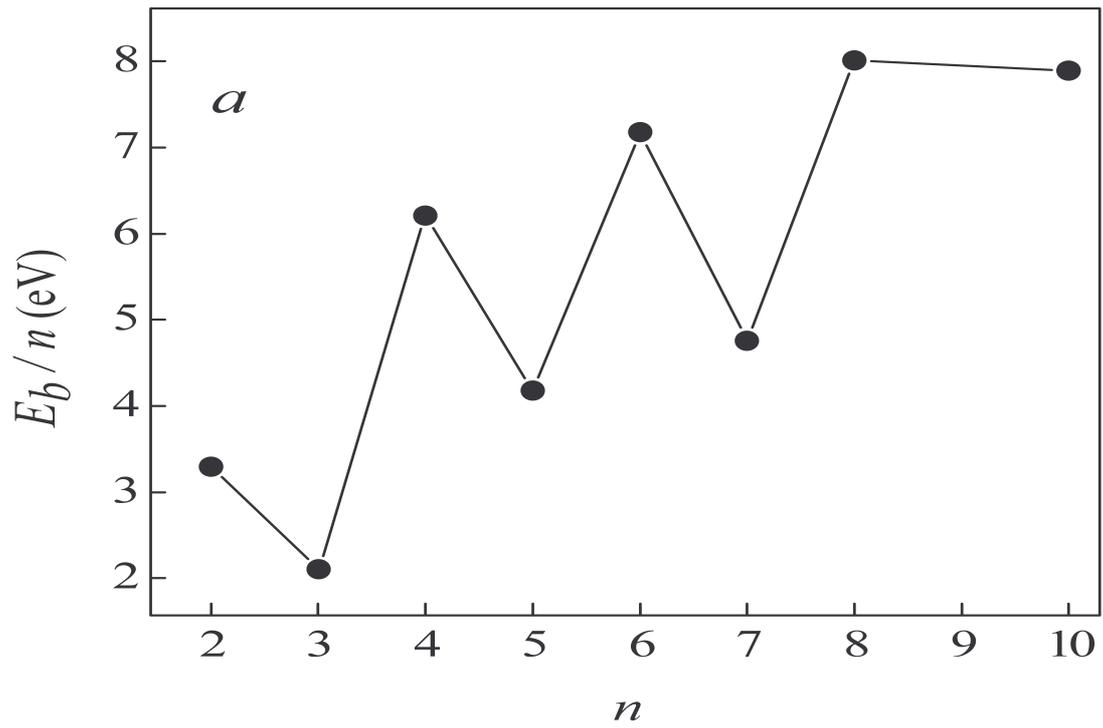

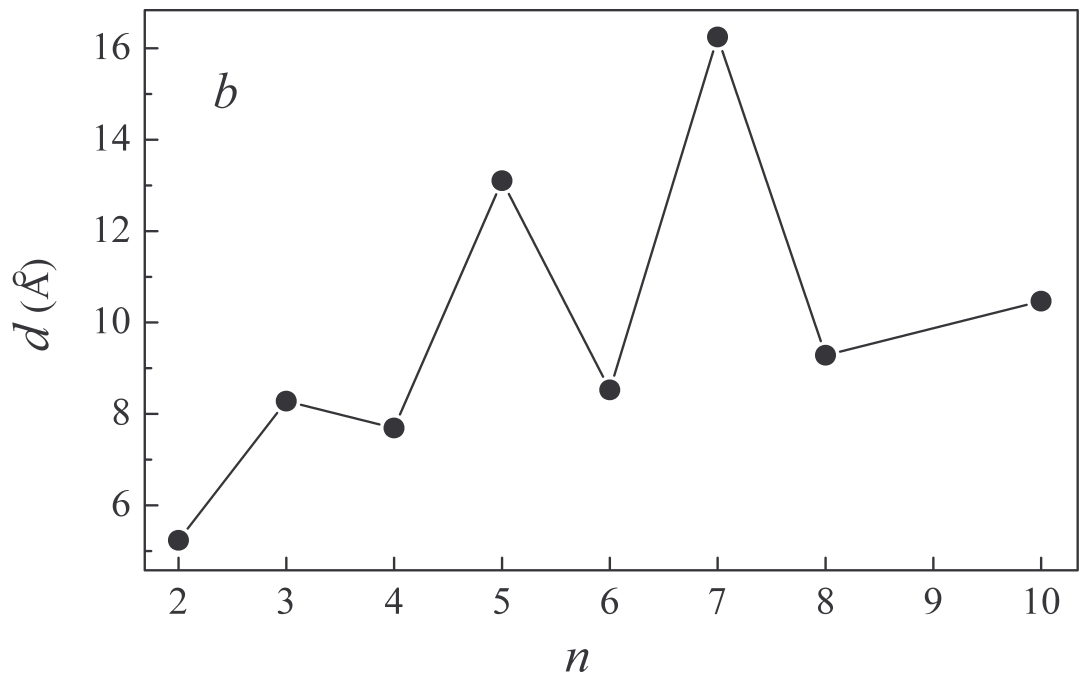

Fig.3. The results of the *ab initio* calculations of $(ZnAs_2)_n$ cluster binding energy per formula unit (*a*) and maximum diameter of cluster (*b*) versus *n*.

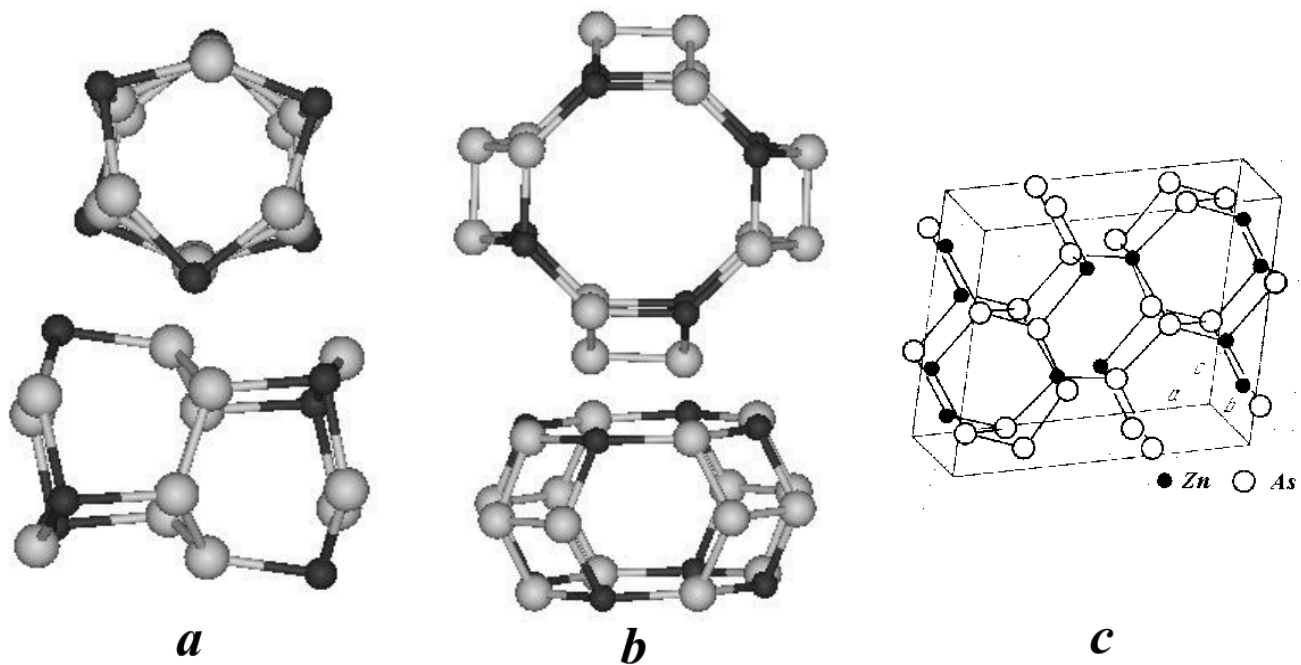

Fig.4. Calculated structure of the most stable ZnAs$_2$ clusters: (*a*) – structure of the (ZnAs$_2$)$_6$ cluster, and (*b*) – structure of the (ZnAs$_2$)$_8$ one, where Zn atoms – black balls, as atoms – grey balls. (*c*) – structure of the unit cell of bulk ZnAs$_2$ crystal (from Ref.[26]).

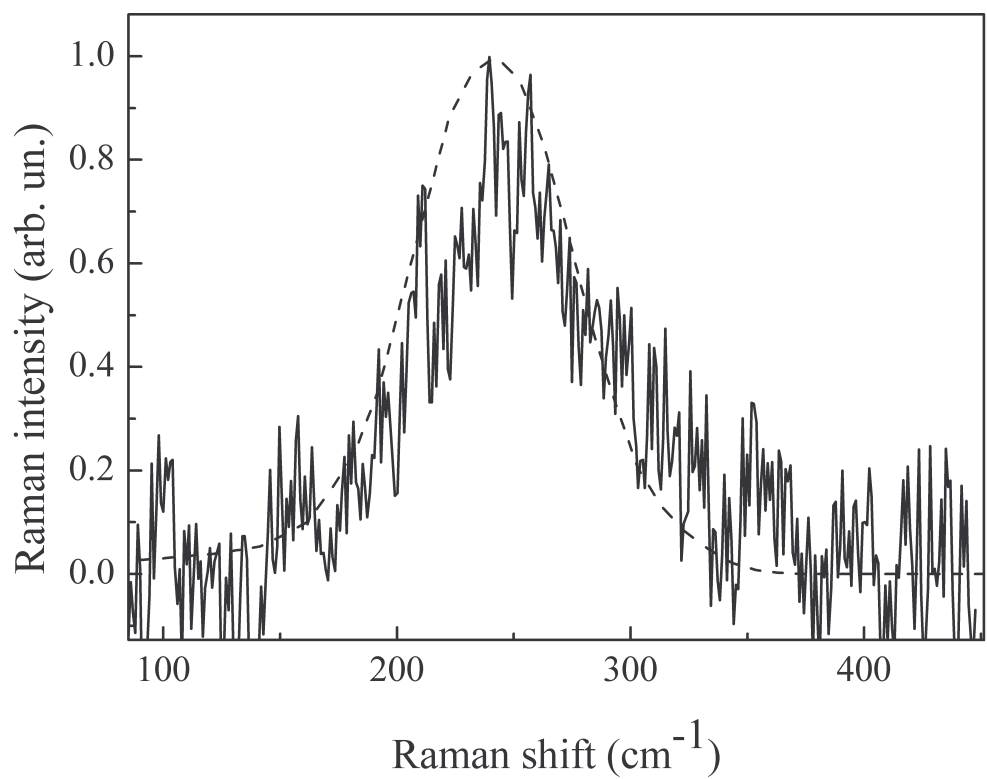

Fig.5. The experimental Raman spectrum of ZnAs$_2$ clusters in zeolite Na-X at the temperature of 293 K (solid line) and the total calculated Raman spectrum of (ZnAs$_2$)$_6$ and (ZnAs$_2$)$_8$ clusters (dashed line).

Table 1. Spectral characteristics of ZnAs$_2$ clusters in zeolite Na-X and produced by laser ablation.

| Spectral position (eV) | | | Blue shift of absorption band (eV) | Stokes shift (eV) | |
| --- | --- | --- | --- | --- | --- |
| Absorption | PL | | | | |
| | Zeolite Na-X | Ablation | | Zeolite Na-X | Ablation |
| 1.919 (B$_1$) | 1.629 (B$_1'$) | 1.309 (B$_2''$) | 0.881 (B$_1$) | 0.290 (B$_1$-B$_1'$) | 0.355 (B$_2$-B$_2''$) |
| 1.664 (B$_2$) | 1.309 (B$_2'$) | | 0.626 (B$_2$) | 0.355 (B$_2$-B$_2'$) | |